\documentclass[a4paper,11pt]{article}
\pdfoutput=1 % if your are submitting a pdflatex (i.e. if you have
\usepackage{jheppub,amsmath} % for details on the use of the package, please
\usepackage[T1]{fontenc} % if needed
\usepackage{bm}% bold math
\usepackage{color}
\usepackage{graphicx}% Include figure files
\allowdisplaybreaks
\usepackage{multirow}

\newcommand\slurp[1]{#1}
\newcommand\sslurp[2]{#1{#2}}
{\catcode`/=\active \expandafter}%
\slurp{\newcommand/}{/\penalty1000\hskip0pt\relax}
{\catcode`:=\active \expandafter}%
\slurp{\newcommand:}{:\penalty1000\hskip0pt\relax}

\newcommand\addspace{\ifcat\nextchar a\spacefactor999. \else.\fi}
{\catcode`\.=\active \expandafter}%
\slurp{\newcommand.}{\futurelet\nextchar\addspace}

\ifx\href\undefined\def\href#1{}\fi
\ifx\texorpdfstring\undefined\def\texorpdfstring#1#2{#1}\fi

\newcommand\myslash{/} \newcommand\mycolon{:}
\newcommand\doi{{\catcode`/=\active \catcode`:=\active \expandafter}\sslurp\realdoi}
{\catcode`/=\active \catcode`:=\active \expandafter}%
\slurp{\newcommand\realdoi[1]{{\let/=\myslash \let:=\mycolon
                               \edef\raw{{http://dx.doi.org/#1}}\expandafter}%
                               \expandafter\href\raw{doi:#1}}}
\newcommand\eprint[2]{{\escapechar-1%
                       \edef\a{\expandafter\string\csname arXiv\endcsname}%
                       \edef\b{\expandafter\string\csname #1\endcsname}%
                       \edef\c{\expandafter\string\csname #2\endcsname}%
                       \edef\d{\noexpand\href{http://arXiv.org/abs/\c}}%
                       \ifx\a\b\expandafter\d\fi{\tt #1:#2}}}
\newcommand{\be}{\begin{equation}}
\newcommand{\ee}{\end{equation}}

\def\d{{\rm d}}
\def\OMIT#1{{}}

\newcommand{\Tr}{\mathrm{Tr}}

\newcommand{\gsim}{\lower.7ex\hbox{$\;\stackrel{\textstyle>}{\sim}\;$}}
\newcommand{\lsim}{\lower.7ex\hbox{$\;\stackrel{\textstyle<}{\sim}\;$}}

\title{\boldmath Dark Gauge Bosons: LHC Signatures of Non-Abelian Kinetic Mixing}

\author[a,b,c]{Carlos A. Arg\"uelles,}
\author[d,e,f]{Xiao-Gang He,}
\author[g]{Grigory Ovanesyan,}
\author[a]{Tao Peng,}
\author[g,h]{and Michael J. Ramsey-Musolf}

\affiliation[a]{Department of Physics, University of Wisconsin, Madison, WI 53706, USA}
\affiliation[b]{Wisconsin IceCube Particle Astrophysics Center, Madison, WI 53703, USA}
\affiliation[c]{Massachusetts Institute of Technology, Cambridge, MA 02139, USA}
\affiliation[d]{INPAC, Department of Physics and Astronomy, Shanghai Jiao Tong University, Shanghai 200240, China}
\affiliation[e]{Department of Physics, National Taiwan University, Taipei 10617, Taiwan}
\affiliation[f]{Physics Division, National Center for Theoretical Sciences, Hsinchu 30013, Taiwan}
\affiliation[g]{Amherst Center for Fundamental Interactions, Department of Physics, University of Massachusetts Amherst, Amherst, MA 01003, USA}
\affiliation[h]{Kellogg Radiation Laboratory, California Institute of Technology, Pasadena, CA 91125, USA}

\emailAdd{caad@mit.edu}
\emailAdd{hexg@phys.ntu.edu.tw}
\emailAdd{ovanesyan@umass.edu}
\emailAdd{mjrm@physics.umass.edu}
\emailAdd{tpeng23@wisc.edu}

\abstract{We consider non-abelian kinetic mixing between the Standard Model SU(2$)_L$ and a dark sector U(1$)^\prime$ gauge group associated with the presence of a scalar SU(2$)_L$ triplet. The magnitude of the resulting dark photon coupling $\epsilon$ is determined by the ratio of the triplet vacuum expectation value, constrained to by $\lsim 4$ GeV by electroweak precision tests, to the scale $\Lambda$ of the effective theory. The corresponding effective operator Wilson coefficient can be $\mathcal{O}(1)$ while accommodating null results for dark photon searches, allowing for a distinctive LHC dark photon phenomenology. After outlining the possible LHC signatures, we illustrate by recasting current ATLAS dark photon results into the non-abelian mixing context.}

\begin{document} 
\maketitle
\flushbottom

\section{Introduction}\label{sec:intro}

%In recent years the models where a Dark Matter (DM) particle is charged under a light dark force have been studied extensively. Typical models assume abelian kinetic mixing between a dark $U(1)'$ sector and the Standard Model (SM) hypercharge \cite{Holdom:1985ag,Foot:1991kb,Fayet:2004bw}. In Refs.\cite{Frandsen:2011cg,Davoudiasl:2012ag,Davoudiasl:2014kua} an additional mass mixing operator between the $Z$ boson and the dark photon is included. For a recent review of the status of the various collider experiments bounds on dark photons see Ref.~\cite{Essig:2013lk} and the references therein. 

%An interesting alternative is the mixing of the non-abelian sector of SM with the abelian dark $U(1)'$. Such mixing assisted with a real Higgs triplet, is the main subject of this paper. In this setup the effective operator that mediates the non-abelian kinetic mixing has dimension five. Such interactions could arise from a UV completion that contains in addition to the real scalar triplet under the SM gauge group,  heavy fermions that are charged under both the SM and the dark $U(1)'$ gauge groups.

%\begin{itemize}
The search for weakly coupled light vector bosons has been a subject of considerable interest in recent years. Searches have been carried out in a number of different contexts, including low energy colliders, meson decays, beam dump experiments, and high-energy colliders (see, {\em e.g.},  Refs.~\cite{Essig:2013lk,Curtin:2014cca} and references therein). Theoretical studies typically assume that interactions of the \lq\lq dark photon\rq\rq\,  with the visible sector are mediated by abelian kinetic mixing between the Standard Model (SM) hypercharge and the dark U(1)$^\prime$ gauge groups \cite{Holdom:1985ag,Foot:1991kb,Fayet:2004bw}. For the \lq\lq dark $Z$\rq\rq, mixing with the SM $Z$-boson may also occur {\em via} the mass terms in the Lagrangian\cite{Davoudiasl:2012ag,Davoudiasl:2014kua}. For both abelian and mass-mixing, the effects arise at the level of renormalizable operators. The resulting coupling of the dark vector bosons to the SM are then parameterized by a dimensionless parameter $\epsilon$ that is constrained by experiment to be $\lsim 10^{-3}$ or smaller when for dark boson masses below $\sim 10 $ GeV. The small scale of $\epsilon$ has no obvious origin in this context, so one must resort to models to explain why it is not $\mathcal{O}(1)$.

In this study, we observe that non-abelian kinetic mixing between the U(1)$^\prime$ and the SM SU(2$)_L$ gauge groups, encoded in non-renormalizable operators,  can provide a simple explanation without assuming tiny operator coefficients in the effective theory. Doing so requires augmenting the SM field content with additional bosons gauge bosons transforming non-trivially under SU(2$)_L$. For concreteness, we consider the scalar triplet\footnote{We list the quantum numbers in the order $SU(3)_\mathrm{C}\times SU(2)_\mathrm{L}\times U(1)_\mathrm{Y}\times G_\mathrm{D}$, where $G_\mathrm{D}$ is the dark gauge group.} $\Sigma\sim (1, 3, 0,0)$ and focus on the dimension-five operator 
\be
\label{eq:dimensionfive}
\mathcal{O}_{WX}^{(5)} = 
-\frac{\beta}{\Lambda}\,\text{Tr}\left(W_{\mu\nu}\Sigma\right) X^{\mu\nu}
\ee
where  $X^{\mu\nu}$ and $W^{\mu\nu}$ are the U(1)$^\prime$ and SU(2$)_L$ field strength tensors, respectively; $\Sigma = \Sigma^a T^a$ with $T^a$ being the SU(2$)_L$ generators; and $\Lambda$ is the mass scale associated with fields that have been integrated out in generating the operator. 
A non-zero vacuum expectation value $\langle \Sigma^0\rangle \equiv v_\Sigma$ will lead to mixing between the U(1)$^\prime$ boson $X_\mu$ and the neutral SU(2$)_L$ gauge boson $W^3_\mu$. The mixing parameter is then given by
\be
\label{eq:dimfiveeps}
\epsilon = {\beta\sin\theta_W}\, \left(\frac{v_\Sigma}{\Lambda}\right)\ \ \ ,
\ee
where $\theta_W$ is the weak mixing angle. For non-vanishing mixing parameter, $X_\mu$ inherits all couplings of the photon to SM fermions but rescaled by the universal factor $\epsilon$, whose magnitude is controlled by 
the scale ratio $v_\Sigma/\Lambda$. Importantly, constraints from electroweak precision tests constrain the triplet vev to be relatively small: $v_\Sigma\lsim 4$ GeV. Thus, $\epsilon$ will satisfy the experimental bounds for $\Lambda$ larger than about one TeV for $\beta\sim \mathcal{O}(1)$. 

The idea of non-abelian kinetic mixing is not original to us. The authors of  Ref.~\cite{Chen:2009dm} considered U(1)$_Y\times$ SU(2$)^\prime$, with the latter factor being a dark SU(2) gauge group~\cite{Ko:2014dm}. Dark SU(2) gauge invariance requires introduction of an additional scalar triplet $\Sigma_D\sim\left(1,1,0,3\right)$, allowing for a dimension five mixing operator analogous to that of Eq.~(\ref{eq:dimensionfive}). In contrast to the present case, however, the dark triplet vev can have any magnitude, and for large values, a small $\epsilon$ requires a commensurately small operator coefficient. In a follow-up work~\cite{Chen:2009ab} applications for astrophysical anomalies and other constraints are studied in this scenario. In Ref.~\cite{Cline:2014kaa} this non-abelian kinetic mixing is used to explain the X-ray line at 3.55 keV.

More recently, the authors of Ref.~\cite{Barello:2015bhq} considered SU(2)$_L\times$U(1)$^\prime$ kinetic mixing {\em via} the dimension six operator 
\be
\label{eq:dimensionsix}
\frac{C}{\Lambda^2}\, H^\dag T^a H W^a_{\mu\nu} X^{\mu\nu}
\ee
where $H$ is the SM Higgs doublet, leading to $\epsilon\sim C (v/\Lambda)^2$. Assuming this operator arises at one-loop, one has $\Lambda\sim 4\pi m_\varphi$, where $m_\varphi$ is the mass of the mediator $\varphi$ in the loop. For $\Lambda\gsim 10$ TeV (or $m_\varphi\gsim 1$ TeV), one may satisfy the experimental constraints on $\epsilon$ for $C\sim \mathcal{O}(1)$. The authors of this work consider an explicit model with a scalar mediator $\varphi\sim(1,3,0,q_D)$ and a dark Higgs $h_D\sim(1,1,0,q_D)$ that is responsible for generating the dark photon mass. A detailed analysis of the collider signatures associated with the dark bosons is given.

In what follows, we concentrate on the collider signatures associated with the dimension five operator (\ref{eq:dimensionfive}) rather than on construction of an explicit dark sector mediator model. In particular, we note that final states containing one or more $X$ bosons may be produced through  two distinct mechanisms, each of which involves $\mathcal{O}_{WX}^{(5)}$ directly: (1) Drell-Yan pair production of $\Sigma$ states,  $pp\rightarrow V\rightarrow \Sigma\Sigma$, followed by the $\mathcal{O}_{WX}^{(5)}$-induced decay $\Sigma\rightarrow XV$, resulting in a $XXVV$ topology;  (2) direct production {\em via} $\mathcal{O}_{WX}^{(5)}$, $pp\rightarrow V^\ast\rightarrow X\Sigma$,  followed by the the decay $\Sigma\rightarrow XV$,  generating a final state of the topology $XXV$. For sufficiently large $\beta/\Lambda$ the direct production mechanism (2) may dominate. In this case, $v_\Sigma$ must be sufficiently small to ensure the experimental constraints on $\epsilon$ are satisfied. Conversely, for smaller $\beta/\Lambda$ (larger $v_\Sigma$ for a given $\epsilon$), production will occur primarily through the Drell-Yan process\footnote{In principle, the same set of possibilities applies to the operator (\ref{eq:dimensionsix}); in practice, they are less likely to be realized, since the minimum value of $\Lambda$ is roughly ten times larger than for the interaction (\ref{eq:dimensionfive}) and since the dimension six operator carries a quadratic dependence on the inverse mass scale. Thus, consideration of the dark sector mediators responsible for (\ref{eq:dimensionsix}) as analyzed in Ref.~\cite{Barello:2015bhq} may be the most promising probe in the latter case.}. For similar reasons, the $\Sigma$-decay branching ratios will also carry a dependence on $\beta/\Lambda$ (and, thus, on $v_\Sigma$ for fixed $\epsilon$). In what follows, we delineate several general parameter space regimes associated with this interplay of parameters. 

For concrete illustration, we then consider the present LHC sensitivity for the regions of parameter space where the direct production mechanism dominates and where the $\Sigma\to VX$ branching ratio is close to unity. For this parameter space region and for $m_X > 2 m_\mu$, one expects displaced vertices associated with $X\to\mu^+\mu^-$ decays, where the dimuon pair appears as a lepton jet. The ATLAS collaboration has performed a search for events of this type that involve two or four lepton jets\cite{Aad:2014yea}. We carry out a simple recast of the corresponding  ATLAS bound on long-lived dark bosons  for our scenario, noting that the ATLAS search is inclusive and  accommodates additional, unobserved, final state SM gauge bosons. For  dark boson mass $m_X$  in the range $0.2\,\text{GeV}\leq m_X\leq  2 \text{\,GeV}$ we find that the present ATLAS exclusion can extend to $\Lambda/\beta\sim$ several hundred GeV, depending on the value of $v_\Sigma$. As we discuss in Section ~\ref{sec:ATLASrecast}, the present reach may lie on the border of the region of validity of the effective theory. Consequently, one should consider our results as indicative of the LHC 8 TeV sensitivity to the parameters of this scenario rather than as quantitatively definitive. We, thus,  also discuss the possibilities for future LHC tests of this scenario that would probe higher mass scales, including searches that would identify the SM final state gauge bosons. 

Our discussion of this scenario and collider analysis is organized as follows. In Section \ref{sec:formalism} we review the setup of the triplet-assisted non-abelian kinetic mixing. In Section \ref{sec:collidersignatures} we outline distinctive LHC signatures for our scenario and in Section~\ref{sec:ATLASrecast} we present the recast of ATLAS bounds on dark photons for the non-abelian kinetic mixing. Finally, we conclude in Section \ref{sec:conclusions}.

\

\section{The Model}\label{sec:formalism}
\begin{figure}
\begin{center}
\includegraphics[width=16cm]{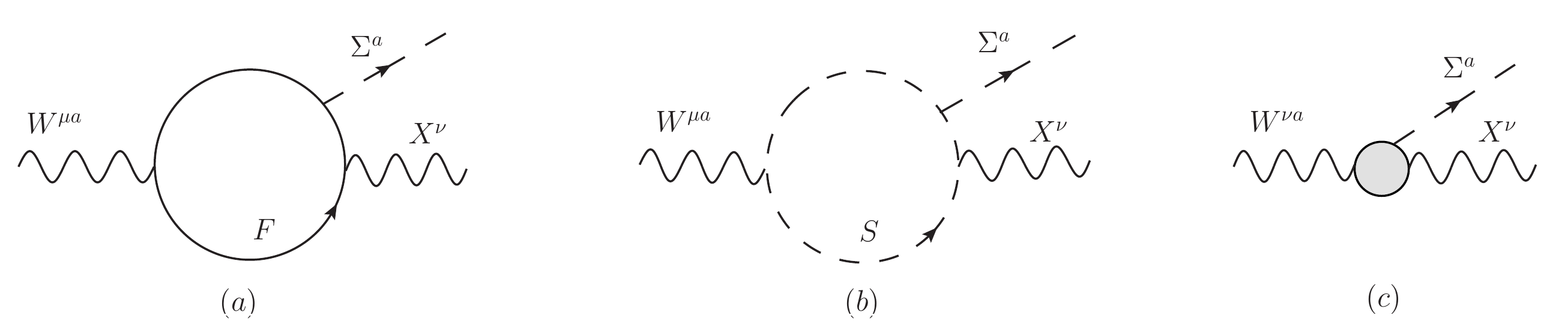}
\caption{Feynman graphs that may generate non-abelian mixing SU(2$)_L\times$ U(1$)^\prime$. Here, the mediators in the loop may be  (a) fermions, (b) scalars, or (c) other degrees of freedom associated with non-perturbative dynamics.}\label{fig:mixinggraphs}
\end{center}
\end{figure}
We add to SM Lagrangian dimension four operators involving dark photon and the real triplet fields, as well as dimension five effective operators:
\begin{eqnarray}
&&\mathcal{L}=\mathcal{L}_{\text{SM}}+\Delta \mathcal L^{(d=4)}+\Delta\mathcal L^{(d=5)}+\dots\,\,.
\end{eqnarray}
The dimension four and five operators we take to be of the form:
\begin{eqnarray}
&&\Delta \mathcal L^{(d=4)}=-\frac{1}{4}X_{\mu\nu}X^{\mu\nu}+\frac{\epsilon_0}{2\,c_W} B_{\mu\nu}X^{\mu\nu}+\Tr\left[\left(D_{\mu}\Sigma\right)^{\dagger}D_{\mu}\Sigma\right]-V(\Sigma,H)+\Delta\mathcal{\tilde{L}}^{(d=4)}\label{eq:BSMlagrangian},\nonumber\\
&&\Delta\mathcal L^{(d=5)}=-\frac{1}{\Lambda}\,\text{Tr}\left(W_{\mu\nu}\Sigma\right) \left(\alpha B^{\mu\nu}+\beta X^{\mu\nu}\right)\equiv \mathcal{O}_{WB}^{(5)}+\mathcal{O}_{WX}^{(5)} \,.\label{eq:newphysicsoperator}
\end{eqnarray}
Here, $\Delta \mathcal L^{(d=4)}$ contains the usual abelian ($XB$) kinetic mixing term and $c_W$ is the cosine of the weak mixing angle. The terms breaking the dark $U(1)'$ gauge group are not explicitly presented and are part of $\Delta\mathcal{\tilde{L}}^{(d=4)}$. The real triplet field $\Sigma$ and the scalar triplet-doublet potential are given by~\cite{FileviezPerez:2008bj}:
\begin{eqnarray}
\Sigma=\frac{1}{2}\left( \begin{array}{cc}
\Sigma^0 & \sqrt{2}\Sigma^+  \\
\sqrt{2}\Sigma^- & -\Sigma^0  \end{array} \right),\qquad D_{\mu}\Sigma=\partial_{\mu}\Sigma+ig \left[\sum_{a=1}^3W_{\mu}^a T^a,\Sigma\right]\,, 
\end{eqnarray}
\begin{eqnarray}
V(H,\Sigma)=-\mu^2 H^{\dagger}H+\lambda_0\left(H^\dagger H\right)^2-\mu_{\Sigma}^2\,G+{b_4}G^2+a_1\,H^{\dagger}\Sigma H+{a_2}H^{\dagger}H G,
\end{eqnarray}
where $G\equiv \text{Tr}\Sigma^\dagger \Sigma=\frac{\left(\Sigma^0\right)^2}{2}+\Sigma^+\Sigma^-$. In the notation of Ref.~\cite{FileviezPerez:2008bj}, $G=F/2$. 

Given a UV complete theory one may integrate out heavy states that have both SM and dark charges, as illustrated in Figure~\ref{fig:mixinggraphs}. We leave the model-dependent details of the full theory unspecified, focusing instead on $\mathcal{O}_{WX}^{(5)}$ and the corresponding collider phenomenology.  
%This is in contrast to Ref.~\cite{Barello:2015bhq} where the full theory has been constructed and the collider phenomenology was focused on the production of the mediator particles and branching ratios involving new states, present in the concrete realization of the model. 
In addition it is possible that similar graphs as in Figure~\ref{fig:mixinggraphs} generate the effective dimension five operator $\mathcal{O}_{WB}^{(5)}$. After electroweak symmetry breaking (EWSB), this operator will contribute to the $S$ parameter:
\begin{eqnarray}
\alpha_{\text{em}} S=4 c_W s_W\,\frac{\alpha v_{\Sigma}}{\Lambda}.
\end{eqnarray}
This sets a $90\%$ CL bound $\alpha v_{\Sigma}/\Lambda\lesssim 0.0008$. We will henceforth set $\alpha=0$ and concentrate on the phenomenology associated with $\mathcal{O}_{WX}^{(5)}$.

Before proceeding, we comment here that kinetic mixing of gauge bosons can also be realized for non-abelian groups. For example, 
for a SU(N)$\times$SU(M) gauge theory with gauge fields $W$ and $Y$, one can introduce 
a scalar field $\Delta^{ab}$ transforming as the adjoint representation under both the SU(N)$\times$SU(M) groups, with indices \lq\lq $a$\rq\rq\,  and \lq\lq $b$\rq\rq\,  corresponding to SU(N) and SU(M), respectively. 
In analogy with $\mathcal{O}_{WX}^{(5)}$, one can construct the  $d=5$ operator $W^{a \mu \nu} Y^b_{\mu\nu} \Delta^{ab}$. 
A non-vanishing vev for $\Delta^{ab}$ will lead to  kinetic mixing between $W$ and $Y$. One may also construct renormalizable models that generate this operator at the one-loop level. We defer a detailed consideration of this possibility to a future study.

\section{Collider phenomenology}\label{sec:collidersignatures}
%\begin{itemize}
%\item{Propose the mechanism for Drell-Yan production of $pp\rightarrow H^{+} H^{-}$ with subsequent decays $H^{\pm}\rightarrow W^{\pm} X$ and $X\rightarrow l^+l^-$ giving a displaced vertex. Using Michael's paper from 2008 and calculating the decay rate of $H^{\pm}\rightarrow W^{\pm} X$ calculate branching of the decays of $H^+, H^-$ to $X$ boson, and estimate needed integrated luminosity to see the signal. Does this particular channel contribute to $W^+ W^-$ $2-\sigma$ access of ATLAS and CMS ? }

%\item{Mention Michael's new mechanism of producing $pp\rightarrow W^+\rightarrow H^+ X$ with subsequent decay of $X$ to leptons with displaced vertex. Production cross section is known, the decay of $W^+$ for heavy $H^+$ is via off-shell $W^+$. Leave analysis for future work.}
%\end{itemize}

\begin{figure}
\begin{center}
\includegraphics[width=3.45cm]{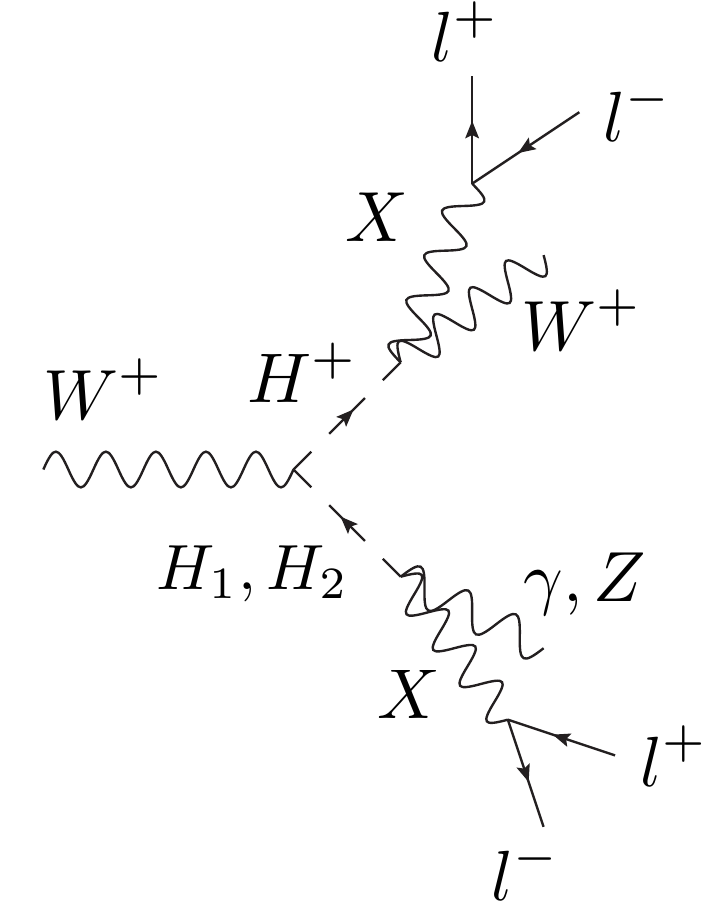}\quad \includegraphics[width=3.45cm]{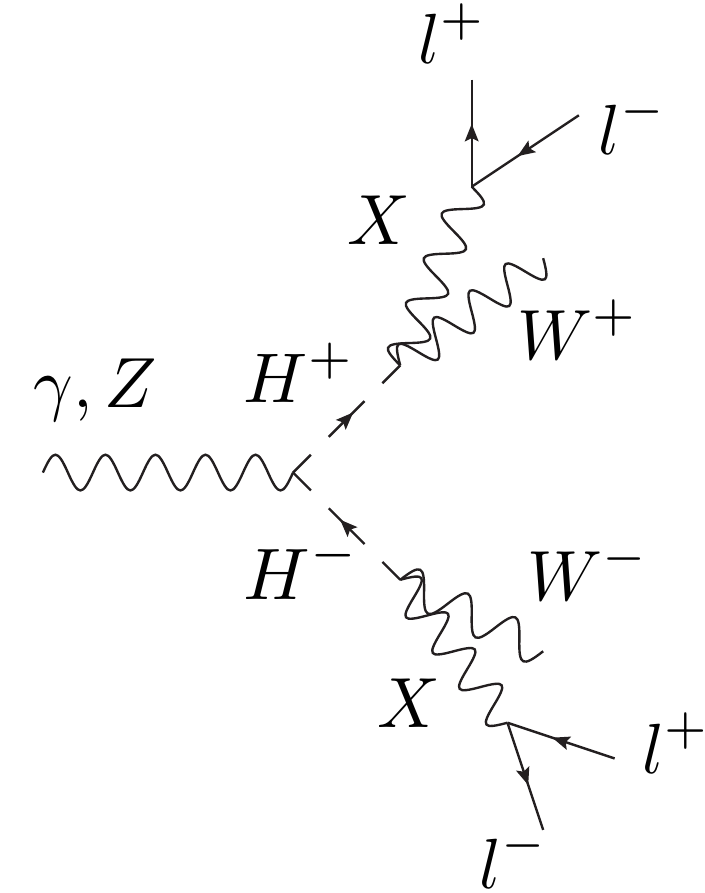}\quad\includegraphics[width=3.45cm]{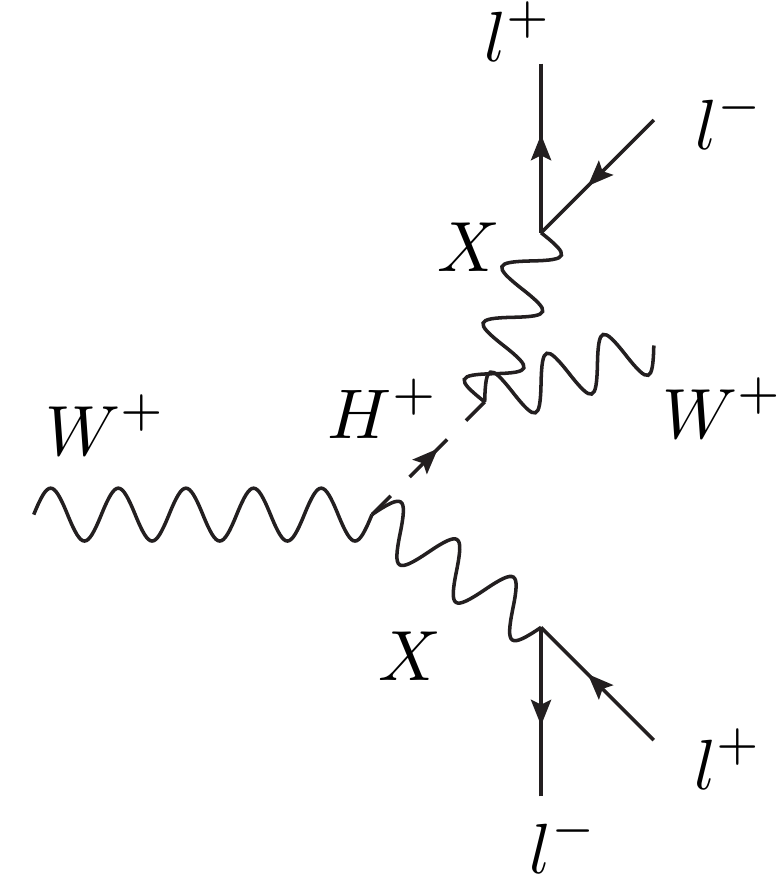}\quad \includegraphics[width=3.45cm]{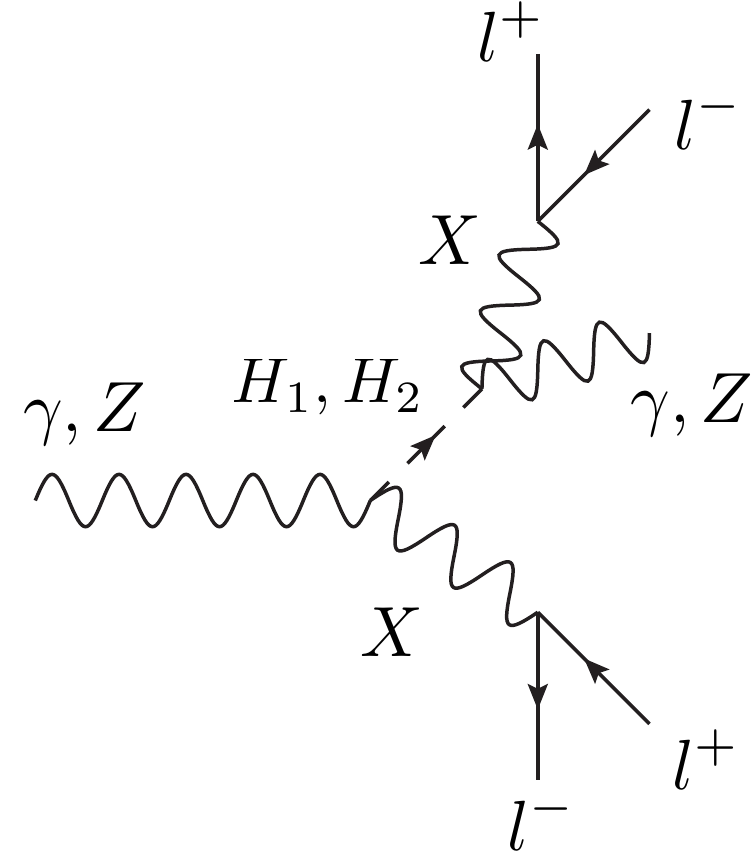}
\put(-385,-20){\text{(a)}} \put(-275,-20){\text{(b)}} \put(-163,-20){\text{(c)}} \put(-55,-20){\text{(d)}}
\caption{Feynman graphs for LHC production and decay of the particles in the triplet-assisted non-abelian mixing model. Diagrams (a,b) indicate scalar pair production, followed by $\mathcal{O}_{WX}^{(5)}$-mediated scalar decays. Diagrams (c,d) inducate $\mathcal{O}_{WX}^{(5)}$-mediated production and decays. In all graphs, the incoming vector boson is virtual.}\label{fig:LHCsignatures}
\end{center}
\end{figure}

%\begin{figure}
%\begin{center}
%\includegraphics[width=8.1cm]{fig1}\qquad \includegraphics[width=8.1cm]{fig2}
%\includegraphics[width=3.45cm]{FeynGraphs2prime}\quad \includegraphics[width=3.45cm]{FeynGraphs1}\quad\includegraphics[width=3.45cm]{FeynGraphs1prime}\quad \includegraphics[width=3.45cm]{FeynGraphs2}\\
%\caption{Feynman graphs for LHC production and decay of the particles in the triplet assisted non-abelian dark force model.}\label{fig:LHCsignatures}
%\end{center}
%\end{figure}
In the presence of $\mathcal{O}_{WX}^{(5)}$, the collider phenomenology associated with the real triplet can differ substantially from what has been considered previously in Ref.~\cite{FileviezPerez:2008bj}. To illustrate the key features, we will make the following  assumptions: 
\begin{itemize}
\item[(a)] The potential parameters are chosen so as to render the doublet-triplet mixing angle -- proportional to $v_\Sigma$ --  to be small, but non-vanishing. In this case the neutral scalar sector will consist of two states, $H_{1,2}$, with $H_1$ being primarily the SM Higgs boson and $H_2$ being primarily $\Sigma^0$. In the charged scalar sector, doublet-triplet mixing implies that the physical charged triplet states $H^\pm$ are not pure triplet states, but rather mixtures of $\Sigma^\pm$ and the charged components of the doublet, with the the other combination providing the longitudinal components of the massive weak gauge bosons. Note that in the absence of doublet-triplet mixing, SU(2$)_L\times$U(1$)_Y$ gauge invariance precludes $\Sigma$ from coupling to the SM fermions. The presence of a non-vanishing mixing angle then introduces a coupling of $H^\pm$, $H_2$ to the SM fermions through the SM Yukawa interactions\footnote{For generic choices of scalar potential parameters, the magnitude of the neutral doublet-triplet mixing angle falls well below the upper bound implied by Higgs-boson signal strengths\cite{Profumo:2014opa}. See Ref.~\cite{FileviezPerez:2008bj} for a detailed analysis of the dependence of the mixing angle on the potential parameters.}.
\item[(b)] For $v_\Sigma = 0$, the triplet states have a common mass, give by  $m_\Sigma^2 = -\mu_\Sigma^2 + a_2 v^2/2$. Electroweak loops raise the mass of the charged components with respect to that of the neutral component by $\sim 166$ MeV, allowing for the decay $H^+\to H_2 \pi^+$. Our choice of the potential parameters will not substantially alter this splitting even for $v_\Sigma\not=0$.
\end{itemize}
With these comments in mind, we now consider the production and decays of the triplet-like scalars.
%LHC signatures of a triplet scalar added to SM, including production and decay, have been explored in detail in Ref.\cite{FileviezPerez:2008bj}. The mass eigenstates of neutral scalar particles in our theory are $H_1, H_2$ which are combinations of $h^0, \sigma^0 (=\Sigma^3)$ the neutral components of SM Higgs doublet and the real triplet respectively. This mixing is given by angle $\theta_0$. Mass eigenstates of  charged scalars $H^{\pm}$  and unphysical Goldstone bosons $G^{\pm}$ are combinations of weak eigenstates $\phi^{\pm}, \Sigma^{\pm}$. This mixing is described by the angle $\theta_{\pm}$. In section II.A of Ref.\cite{FileviezPerez:2008bj} all relevant formulas and definitions of angles in terms of short distance parameters of the theory are given. Note that it was found from electroweak precision parameters that mixing angles $\theta_0, \theta_{\pm}$ are tiny. In this limit we have decoupling $H_1\approx h_0, H_2\approx \sigma_0, H^{\pm}\approx \Sigma^{\pm}, G^{\pm}\approx \phi^{\pm}$. All the Feynman rules needed for production of scalar particles $H_1, H_2, H^{\pm}$ at colliders are listed in Appendix C of Ref.~\cite{FileviezPerez:2008bj}. In the Appendix~\ref{sec:AppendixC} below we provide additional Feynman rules relevant for our scenario of non-abelian mixing $XW\Sigma$ and production and decay of $X, H_i$. In that table all momentum of particles flows into the vertex, the polarization and momentum of the dark photon $X$ are $\mu, p$ and that of the SM gauge boson are $\nu, p'$.
\subsection{Production}
\begin{figure}
\begin{center}
\includegraphics[width=7.35cm]{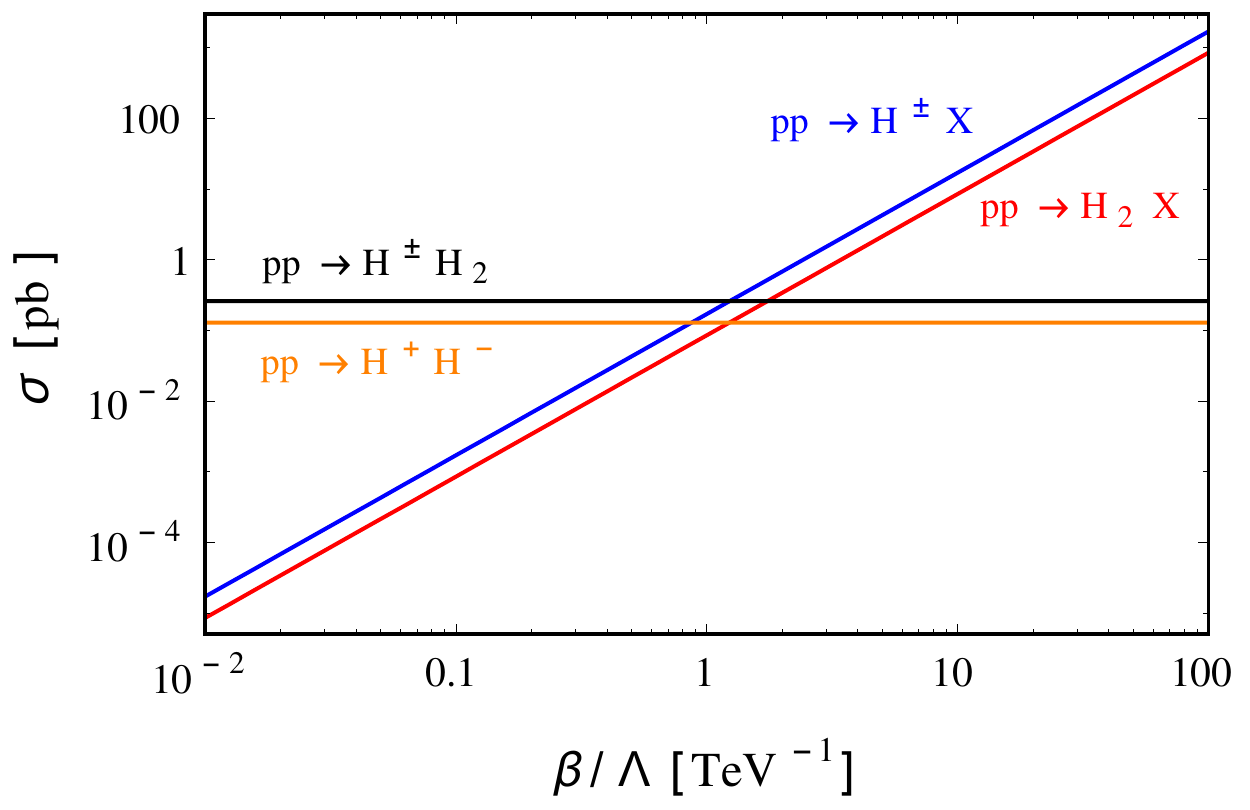}\quad \includegraphics[width=7.35cm]{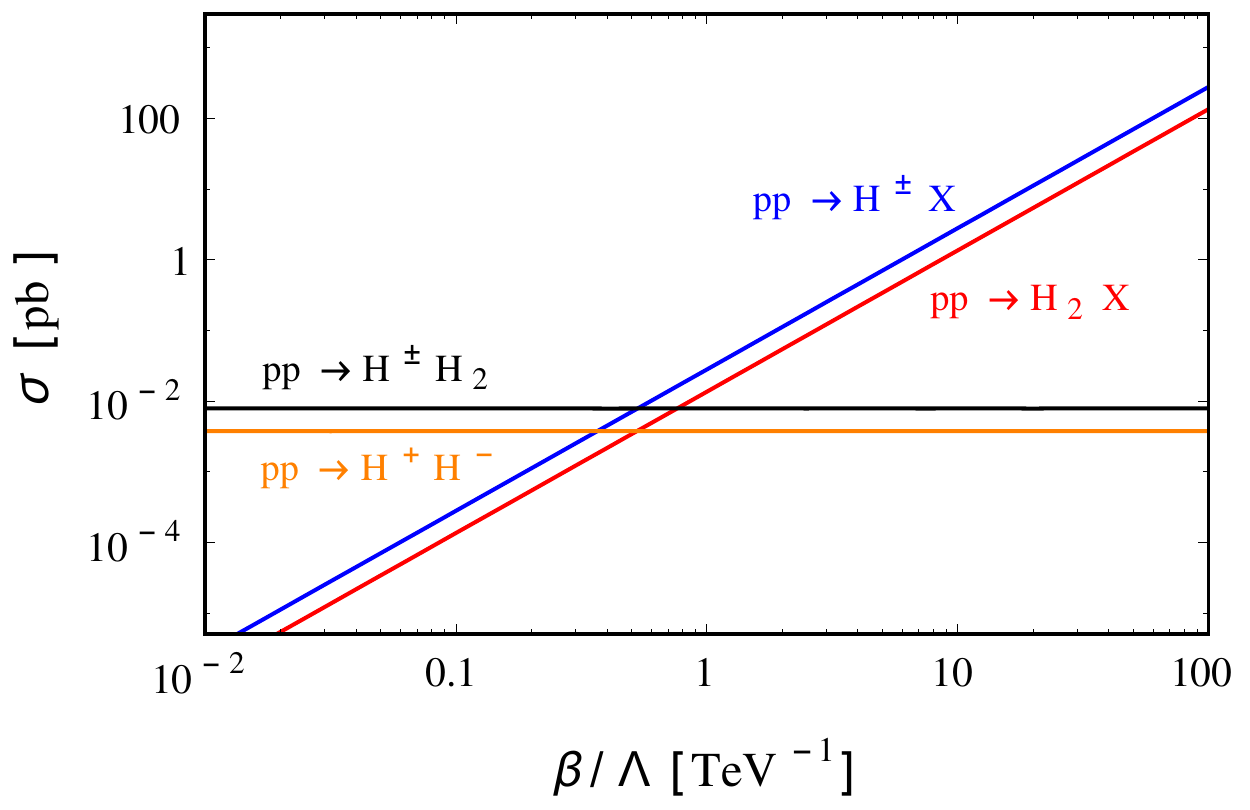}
\caption{Production cross sections for $pp\rightarrow V\rightarrow \phi\phi$ and $pp\rightarrow V\rightarrow X\phi$ for associated triplet-like states $\phi=H^+,H_2$ and a dark photon $X$ with mass $m_X=0.4$ GeV. For the final states containing a single charged scalar and one neutral boson, we have summed the cross sections for both charges [{\em e.g.} $\sigma(H^+H_2)+\sigma(H^-H_2)$]. The left and right panels correspond to $m_\phi=130$ GeV and $m_\phi=300$ GeV, respectively.}\label{fig:productionxsections}
\end{center}
\end{figure}

The LHC production and decay mechanisms of interest are shown in Figure~\ref{fig:LHCsignatures}. Graphs (a) and (b) indicate Drell-Yan pair production, $pp\rightarrow V^\ast\rightarrow \phi\phi$, where $\phi$ denotes any of the physical scalars, with the subsequent decays $\phi\rightarrow XV$, leading to the topology $XXVV$. As discussed above, the $\phi$ states will be predominantly triplet-like.
%{\color{blue}(Need to converge on a unified notation, either we use $V\rightarrow HH$ or $V\rightarrow \Sigma\Sigma$ throughout, I personally am ok with either one.)} 
Graphs (c) and (d) show the $\mathcal{O}_{WX}^{(5)}$-mediated production $pp\rightarrow V^\ast\rightarrow \phi X$, with a subsequent decay $\phi\rightarrow XV$, leading to the topology $XXV$. (Feynman rules for the vertices in Figure~\ref{fig:LHCsignatures} are listed in the Appendix~\ref{sec:AppendixC}.)

In Figure~\ref{fig:productionxsections} we show the the LHC production cross sections for different channels at $\sqrt{s}=8$ TeV . The left panel corresponds to
%\footnote{Note that this is not the mass of the SM Higgs boson, but instead the triplet associated physical neutral and charged Higgs bosons mass: $m_H\equiv m_{H^+}=m_{H_2}$.} 
$m_\phi=130\,\text{GeV}$ and the right one corresponds to $m_\phi=300\text{\,GeV}$. For both masses we observe that for $\beta/\Lambda \lesssim 1\, /\text{\,TeV}$ the Drell-Yan pair production dominates, while for $\beta/\Lambda \gtrsim 1\,/\text{\,TeV}$  $\mathcal{O}_{WX}^{(5)}$-mediated production is the dominant mechanism. For $\sqrt{s}=14$ TeV the corresponding transition between Drell-Yan and $\mathcal{O}_{WX}^{(5)}$-mediated production occurs for approximately the same value of $\beta/\Lambda$.

\subsection{Triplet-like scalar decay branching ratios}
\begin{figure}
\begin{center}
\includegraphics[width=7.3cm]{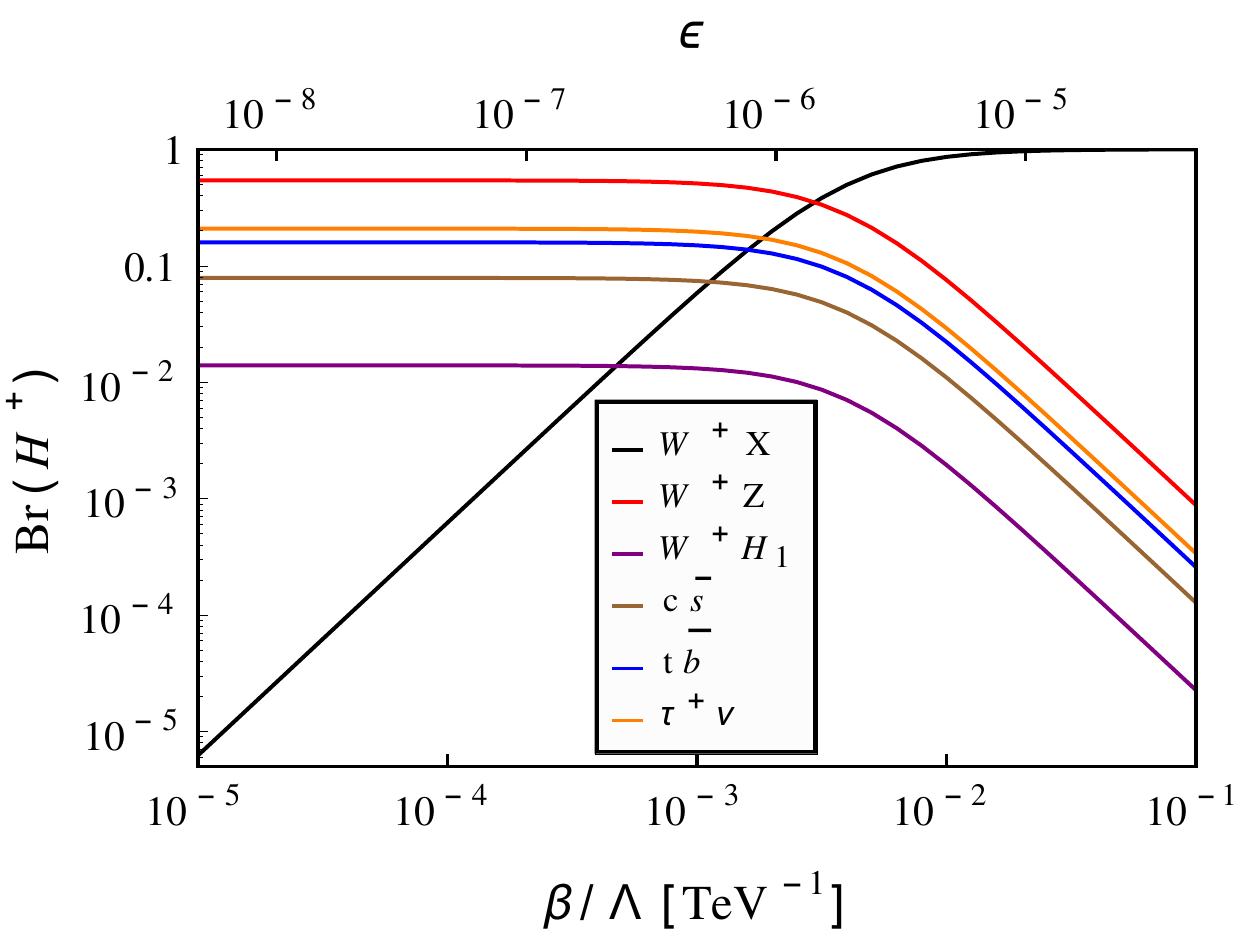}\includegraphics[width=7.3cm]{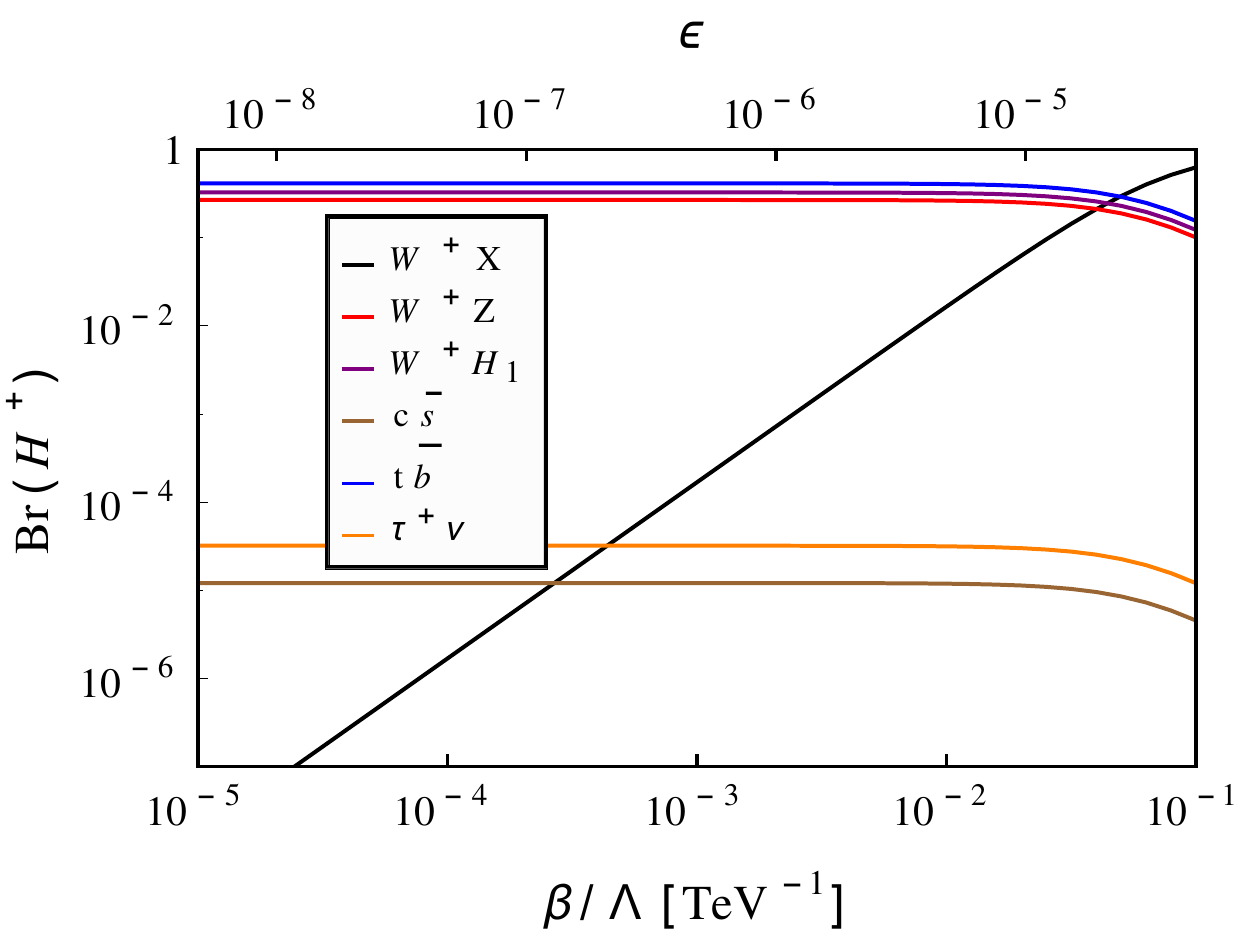}\\
\includegraphics[width=7.3cm]{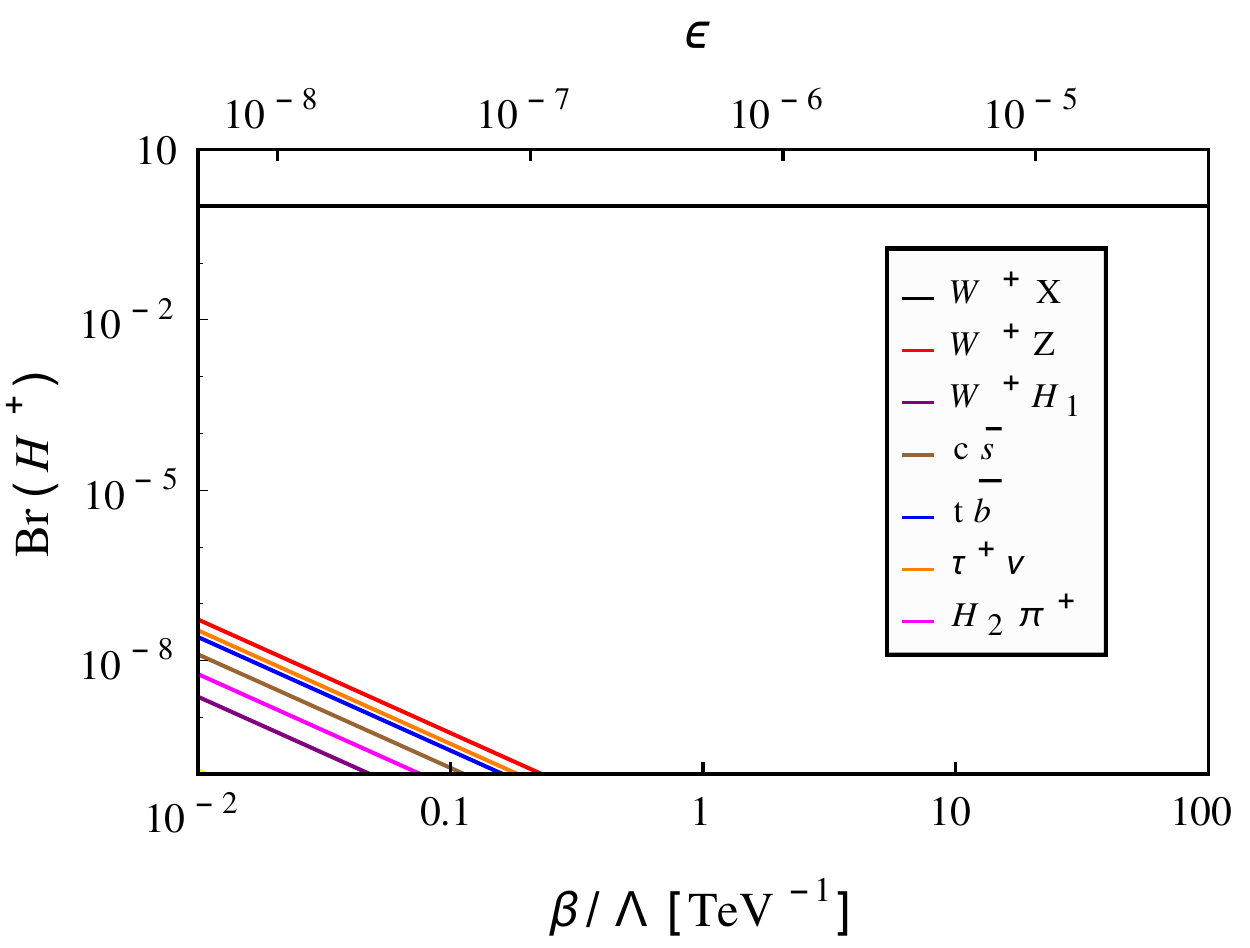}\includegraphics[width=7.3cm]{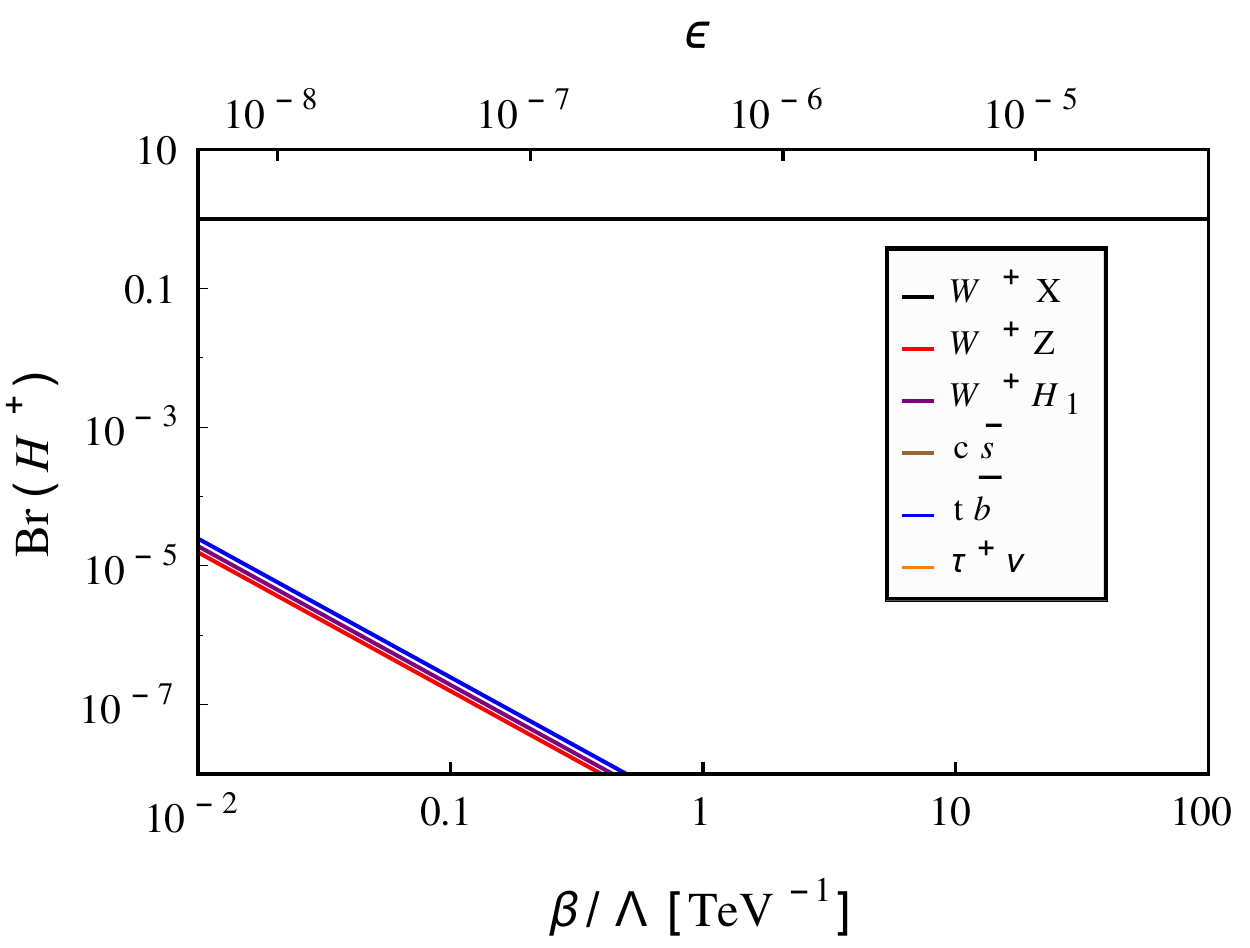}
\caption{Branching ratios for  $H^+$ decays as a function of $\beta/\Lambda$ (bottom horizontal axis) and $\epsilon$ (upper horizontal axis) for $m_X=0.4$ GeV. The top (bottom) row corresponds to $v_\Sigma=1$ GeV ($v_\Sigma=10^{-3}$ GeV), while the left (right) column corresponds to $m_{H^+}=130$ GeV ( $m_{H^+}=300$ GeV). The solid black line indicates the branching ratio for  $H^+\rightarrow W^+ X$. Branching ratios for other final states are as indicated by the legend insert.}\label{fig:Xdecay}
\end{center}
\end{figure}

%\begin{figure}
%\begin{center}
%%\includegraphics[width=8.1cm]{fig1}\qquad \includegraphics[width=8.1cm]{fig2}
%\includegraphics[width=7.35cm]{njet.pdf}\quad \includegraphics[width=7.35cm]{nmuon.pdf}\\
%\includegraphics[width=7.35cm]{ht.pdf}\quad \includegraphics[width=7.35cm]{leadingjetpt.pdf}\\
%\includegraphics[width=7.35cm]{met.pdf}
%\caption{Signal versus a representative $ZZ$ background distributions for different kinematic variables. See text for more details.}\label{fig:differentialdistributionssignalvsbackground}
%\end{center}
%\end{figure}
The triplet-like scalars $H^\pm$ and $H_2$ will decay to $W^\pm X$ and $Z/\gamma\, X$ respectively as well as to other final states as considered in Ref.\cite{FileviezPerez:2008bj}. For illustrative purposes we show the decay width for $H^\pm~\rightarrow~W^\pm X$, which is sufficient for the analysis that we consider  below. The tree level  $H^{\pm}\rightarrow W^{\pm} X$ decay rate is given by 
\begin{eqnarray}
&&\Gamma(H^{\pm}\rightarrow W^{\pm} X)\\
&&=\frac{\sqrt{1-\frac{2(m_X^2+M^2_{W^{\pm}})}{M^2_{H^\pm}}+\frac{(m_X^2-M^2_{W^{\pm}})^2}{M^4_{H^{\pm}}}}}{16\pi M_{H^+}}\left[{\frac{1}{2}\left(M^2_{H^{\pm}}-m_X^2-M^2_{W^{\pm}}\right)^2+M_{X}^2 M_{W^\pm}^2}\right]\,\frac{\beta^2}{\Lambda^2}c_{\mp}^2\,,\nonumber
\end{eqnarray}
where $c_{\mp}$ is the mixing angle associated with diagonalizing the charged scalar sector.
Combined with the other $H^+$ decay channels~\cite{FileviezPerez:2008bj} we compute the branching ratios shown in  Figure~\ref{fig:Xdecay}.  The left and right panels correspond to $m_{H^+}=130\,\text{GeV}$ and $m_{H^+}=300\,\text{GeV}$, respectively. The top panels correspond to $v_{\Sigma}=1\,\text{GeV}$ and the bottom ones to $v_{\Sigma}=1\,\text{MeV}$.

From the plots in Figure~\ref{fig:Xdecay} we see  that for  $v_{\Sigma}=1\,\text{GeV}$, a value near the maximum allowed by electroweak precision tests,  the branching ratio for $H^+\rightarrow W^+ X$ is essentially $100\%$ when $\epsilon\gtrsim 10^{-4}$. For the smaller value of $v_{\Sigma}=1\,\text{MeV}$, the branching ratio is essentially 100\% for all values of $\epsilon$. This translates into the range $\beta/\Lambda\gtrsim 0.1/\text{TeV}$ for the branching ratio to be essentially $100\%$ independent on the value of the vev. For lower values of $\beta/\Lambda$ any branching ratio from zero to one is possible, and the precise value depends strongly on the value  $v_{\Sigma}$.

\subsection{Various regimes for collider phenomenology}
From the foregoing discussion of production and decays, the LHC signatures and detection strategies will vary according to the value of $\beta/\Lambda$. We delineate three regimes leading to distinctive phenomenology for 8 TeV $pp$ center of mass energy:
\begin{itemize}
\item[(1)] {$\beta/\Lambda\sim 1/\text{TeV}$. In this regime we see that  Drell-Yan pair  production  $p\rightarrow \phi\phi$ dominates  In addition the branching ratio for $\phi\rightarrow XV$ decay is close to hundred percent.}
\item[(2)] {$\beta/\Lambda\lsim 0.1/\text{TeV}$. In this regime Drell-Yan pair production remains the dominant mechanism. However, $\text{BR}(\phi\to XV)$ can range from zero to one, depending on value of $v_{\Sigma}$.}
\item[(3)] {$\beta/\Lambda\gtrsim 1/\text{TeV}$. In this regime the $\mathcal{O}_{WX}^{(5)}$-mediated process $pp\rightarrow X \phi$ is  dominant and $\text{BR}(\phi\rightarrow XV)$ is close to one. In this case, the possible final states are indicated in 
Figure~\ref{fig:LHCsignatures}(c,d).}
\end{itemize}
(Recall that for $\sqrt{s} = 14$ TeV, the transition between $\mathcal{O}_{WX}^{(5)}$-mediated production and Drell-Yan pair production also occurs for $\beta/\Lambda
\sim 1/\mathrm{TeV}$.) While all three possibilities above are worth exploration in future, for illustrative purposes we focus here on the third regime.

%In Figure~\ref{fig:differentialdistributionssignalvsbackground} we compare signal kinematic distributions with similar distributions for the $ZZ$ background that was chosen for illustrative purposes. A more complete study with an array of possible backgrounds is needed. We used MadGraph+Pythia+DELPHES {\bf{(NOTE: put in the latest references)}} to simulate events. {\bf{TODO: explain each observable and what values of $\beta/\Lambda$ have been used in the plots.}} What we see from these distributions is that these kinematic variables are effective in separating the signal from the background. 

%\cite{Aad:2014yea,Curtin:2014cca}

%\begin{figure}
%\begin{center}
%\includegraphics[width=8.1cm]{fig1}\qquad \includegraphics[width=8.1cm]{fig2}
%\includegraphics[width=7.2cm]{Hp-decay-no-X.pdf}\quad \includegraphics[width=7.2cm]{Hp-decay-with-X.pdf}\\
%\caption{Branching ratios of the $H^+$ decays in the scenario of a pure triplet (left) and in the model of non-abelian kinetic mixing (right).}\label{fig:brenchingratios}
%\end{center}
%\end{figure}

\section{ATLAS recast}\label{sec:ATLASrecast}
Considering now regime (3), we recast the ATLAS  dark photon search results \cite{Aad:2014yea} into constraints on our scenario. The analysis of Ref.~\cite{Aad:2014yea} assumes the presence of a SM Higgs boson decaying to two new states that radiate two (or four) dark photons, leading to displaced vertices and lepton jets. This is to be compared to our production scenario mediated by an off-shell vector boson $V^\ast$, 
%(for either Drell-Yan or the $\mathcal{O}_{WX}^{(5)}$-mediated production), 
leading to a final state containing two $X$ bosons and an on-shell $V$. Note, that the ATLAS study~\cite{Aad:2014yea} only applied cuts to isolate events with lepton jets and displaced vertices.  No reconstruction of the Higgs boson invariant mass was preformed, nor were cuts on the missing energy applied. Thus, although the ATLAS study was carried out assuming different underlying $X$-boson production 
dynamics, the analysis is sufficiently inclusive to accommodate the scenario considered here as well. Looking to the future, we note that one could likely improve the LHC sensitivity  to  $\mathcal{O}_{WX}^{(5)}$ by including additional criteria needed to identify the final state $V$. 

\begin{figure}[!t]
\begin{center}
\includegraphics[width=6.8cm]{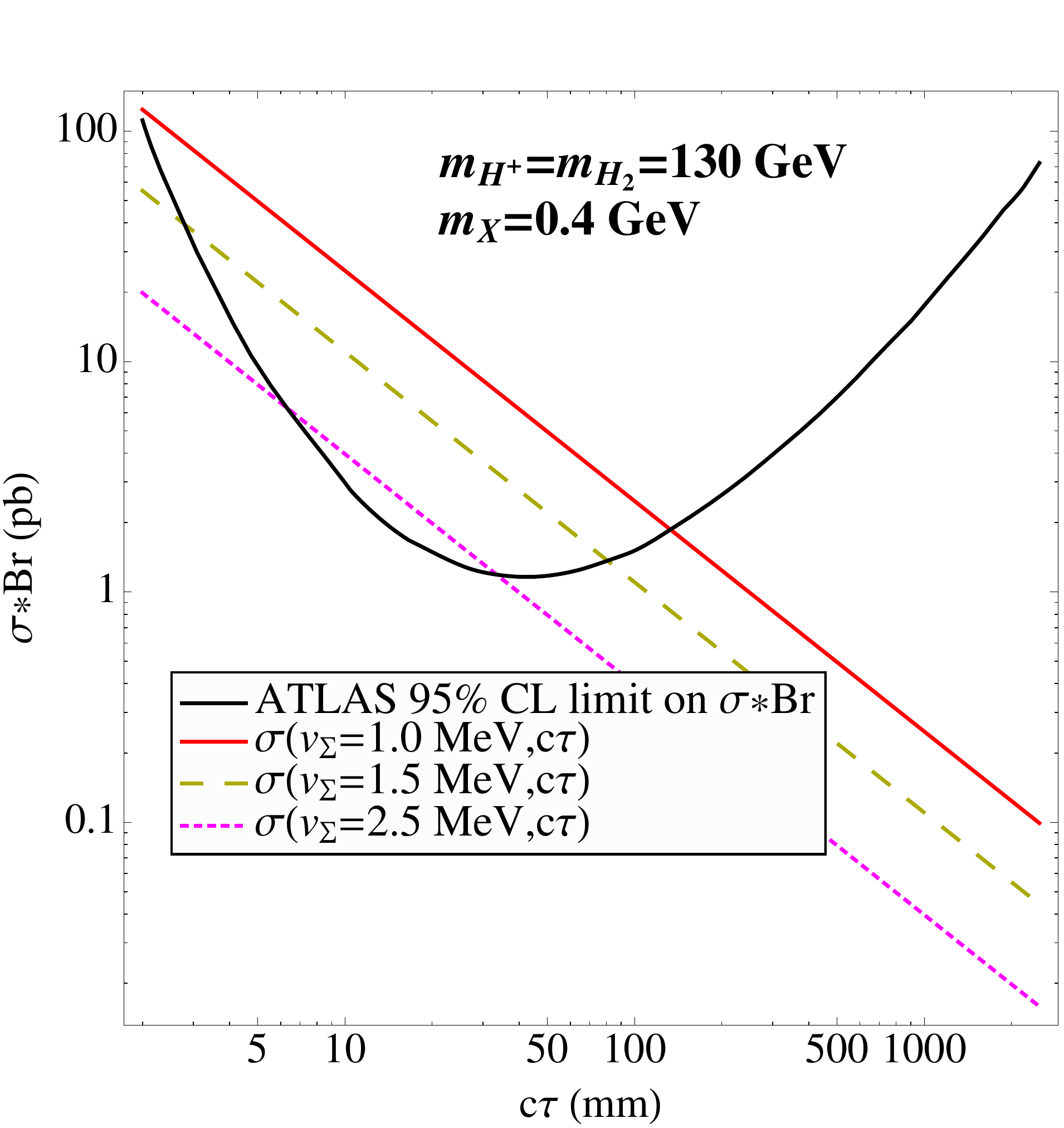}\qquad \includegraphics[width=6.8cm]{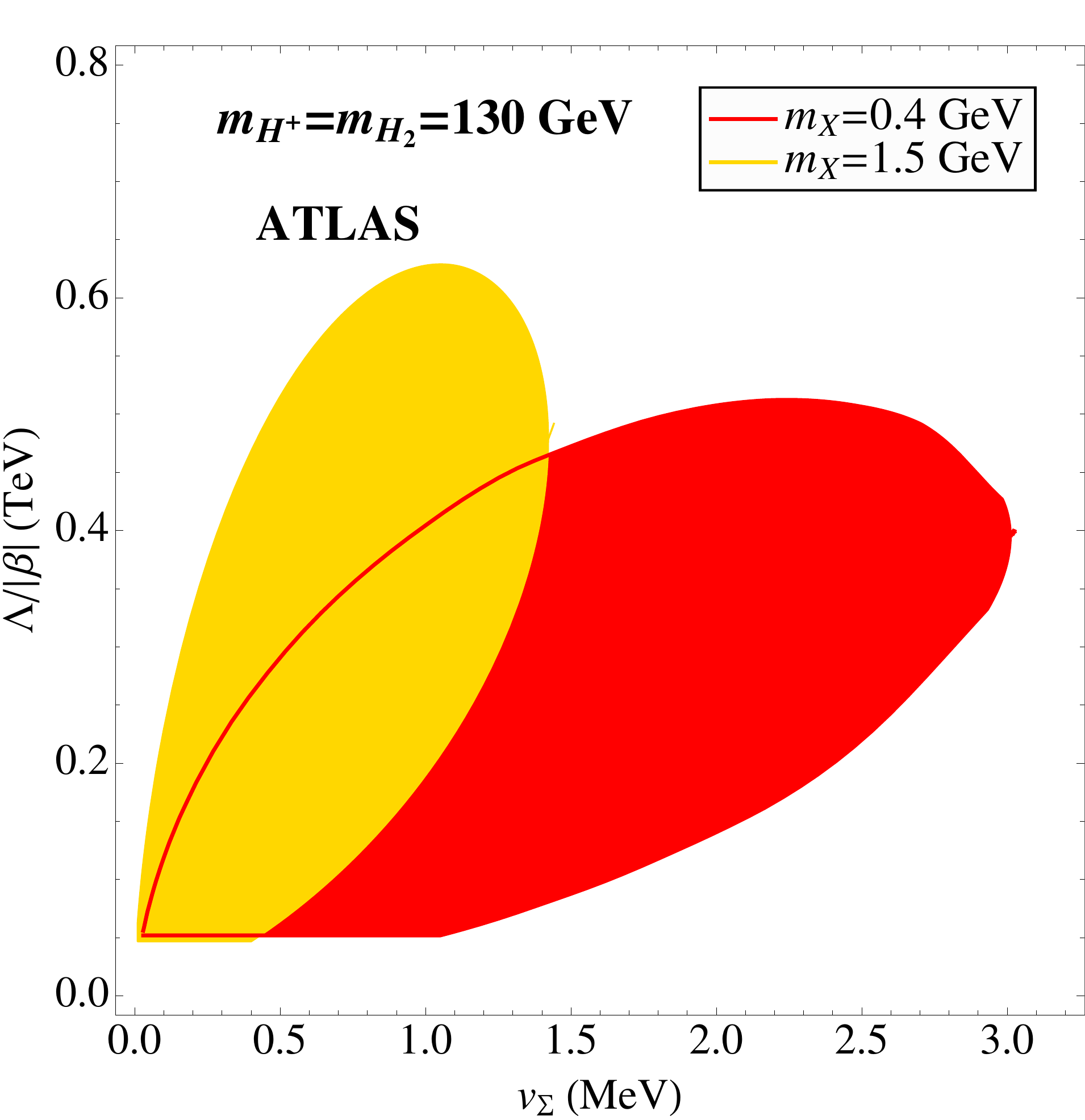}
\caption{Constrains on triplet-assisted non-abelian kinetic mixing, recast from the ATLAS search reported Ref.~\cite{Aad:2014yea}. The left panel gives the exclusion in the ($c\tau$, $\sigma\times\text{BR}$) plane, where the region above the parabola is excluded. The diagonal lines indicate the dependence of $\sigma\times\text{BR}$ on $c\tau$ for different representative choices of $v_\Sigma$. The right panel gives the exclusion region in the($v_\Sigma$, $\Lambda/\beta$) plane for $m_X=0.4$ GeV (red region) and $m_X=1.5$ GeV (yellow region). }\label{fig:ATLASbounds}
\end{center}
\end{figure}
We then translate the ATLAS bounds on $\epsilon-m_X$ parameter space\cite{Aad:2014yea} to the parameter space relevant to our scenario. Certain distinctions between the analysis of Ref.~\cite{Aad:2014yea} and that for our scenario have to be accounted for properly. Specifically, Ref.~\cite{Aad:2014yea}  presents the 95\% C.L. exclusion plots for the signal cross section $\sigma(H)\times \text{Br}(H\rightarrow 2X+\cdots)$ as a function of the dark photon lifetime $c\tau$ ( see the left panel of Figure~16 in that work\footnote{This bound is obtained by excluding from the analysis TYPE2-TYPE2 events, which correspond to both dark photons decaying to jets. This leads to a stronger bound due to corresponding backgrounds.}). In our case, the 95\% C.L. bound applies to $\sigma(\phi X)\times \text{Br}(\phi \rightarrow VX)$. In addition, $\sigma(H)$ and  $\text{Br}(H\rightarrow 2X+\cdots)$ are independent of $\epsilon$ (the dependence on $m_X$ is negligible for very light dark bosons).  The production cross section and branching ratios for our scenario, on the other hand, depend on various combinations of the parameters that govern $\epsilon$, {\em viz}, $\sigma(HX)\sim (\beta/\Lambda)^2\sim 1/(\tau v^2_{\Sigma})$, where $\tau$ is the $X$ lifetime. In making the translation from Ref.~\cite{Aad:2014yea}    we then use the relation in Eq.~(\ref{eq:dimfiveeps}).

%While we will adopt the left panel of Figure~16 as a 95\% bound on the non-abelian kinetic mixing and substitute $\sigma(H)$ with $\sigma(HX)$ and the branching ratio with $\text{Br}(H\rightarrow WX)$, we have to take into account that unlike $\sigma(H), \text{Br}(H\rightarrow 2\gamma_d+X)$ which are independent of the dark photon parameters $\epsilon, m_X$, we have the following explicit dependencies $\sigma(HX)\sim (\beta/\Lambda)^2$ and $\text{Br}(H\rightarrow WX)$ depends on $m_X, \beta/\Lambda, v_{\Sigma}$. As a result, $\sigma(HX)\times\text{Br}(H\rightarrow WX)\sim (\beta/\Lambda)^2\sim \epsilon^2/v_{\Sigma}^2\sim 1/(\tau v^2_{\Sigma})$, and also depends on the dark photon mass. 

%We have used the following relation among  the parameters $\epsilon, \beta/\Lambda, v_{\Sigma}$:
%\begin{eqnarray}
%\frac{\Lambda}{|\beta|}=\frac{s_W\,v_{\Sigma}}{\sqrt{2}|\epsilon|}\,.
%\end{eqnarray}

In the left panel of Figure~\ref{fig:ATLASbounds} we show the ATLAS 95\% CL limit on $\sigma(\phi X)\times \text{Br}(\phi \rightarrow VX)$, summing over all $\phi$, for $m_X=0.4\text{\,GeV}$ (solid black) and  lines of  constant cross section $\sigma(pp\rightarrow \phi X)$ (again, summed over all $\phi$) for three representative values of $v_\Sigma$: $v_{\Sigma}=1\,\text{MeV}$ (solid red), $v_{\Sigma}=1.5\,\text{MeV}$ (dashed olive) and $v_{\Sigma}=2.5\,\text{MeV}$ (dotted magenta). In each case,  $\text{Br}\left(\phi\rightarrow V X\right)\approx 100\%$. For  each line of constant  $v_{\Sigma}$ the points of intersection  with the solid black curve determine the boundaries of the region of excluded $c\tau$. We observe that the ATLAS exclusion then applies to $v_\Sigma$ in the MeV range, well below the $\rho$-parameter bound. These results, together with Eq.~(\ref{eq:dimfiveeps}), lead to constraints in the ($v_\Sigma$, $\beta/\Lambda)$ plane, shown in the right panel of Fig.~\ref{fig:ATLASbounds}. For illustration we consider this translation for  two values of $m_X$: 0.4 GeV (red) and 1.5 GeV (gold). We observe that the exclusion can reach $\Lambda/\beta$ up to several hundred GeV, depending on the value of $m_X$ and $v_\Sigma$. Note that for fixed $\Lambda/\beta$ (fixed $\sigma\times\text{BR}$) , $c\tau$ ($\epsilon$) increases (decreases) with decreasing $v_\Sigma$. Thus, for a given $\Lambda/\beta$ and sufficiently small $v_\Sigma$ (equivalently $\epsilon$), $\tau$ will fall below the ATLAS exclusion curve in the left panel of Fig.~\ref{fig:ATLASbounds}; hence, the exclusion limits on $\Lambda/\beta$ in the right panel weaken with decreasing $v_\Sigma$. 

The foregoing illustrative analysis has endeavored to remain as model-independent as possible. Nevertheless, it is interesting to consider briefly the possible dynamics that may generate $\mathcal{O}_{WX}^{(5)}$ and the corresponding implications for the interpretation of present and prospective LHC results. Figure \ref{fig:mixinggraphs} indicates a few of the possibilities: (a) loops involving new vector-like fermions; (b) loops involving new scalars; (c) non-perturbative dynamics. We comment on the first two possibilities. Considering new vector-like fermions $F$ with mass $M_F$, na\"ive dimensional analysis suggests that $\Lambda/\beta\sim 16\pi^2 M_F/y$, where $y$ is the ${\bar F}F\Sigma$ coupling and where we take the gauge couplings to be $\mathcal{O}(1)$. Since $F$ carries SU(2) charge, it would likely have been observed if sufficiently light. For example, the non-observation of pairs of new charged particles ({\em e.g.}, vector-like leptons) at LHC~\cite{Kumar:2015tna}, may imply a lower bound of 200 GeV $\lsim M_F$ in some cases\footnote{We thank S. Martin for useful discussion of the assumptions underlying the work of Ref. [15].}, implying $\Lambda/\beta \gsim 3.2$ TeV for $y\sim\mathcal{O}(1)$. Significantly larger integrated luminosity and/or a search that exploits final state gauge boson reconstruction would be needed to reach this level of sensitivity. For new electroweak scalars $S$ with mass $M_S$, one has $\Lambda/\beta\sim 16\pi^2 M_S^2F_S/a_S$, where $a_S$ is the $SS\Sigma$ coupling with dimensions of mass, and $F_S$ will depend in part on the $SU(2)$ representation of $S$. Assuming $M_S \gsim 100$ GeV in order to evade LEP II limits and taking $a_S\sim M_S$, and taking $F_S$ to be of order $\mathcal{O}(1)$, it also gives $\Lambda/\beta\sim 1.6$ TeV. However, nothing precludes $a_S$ from being a few times larger than $M_S$, so it is not unreasonable to anticipate $\Lambda/\beta$ being to the upper end of the exclusion region in Fig.~\ref{fig:ATLASbounds}. 

It is possible for the charged triplet state $H^+$ to decay to a pair of mediators at tree-level. In order for this decay to occur, the mediator mass must by less than half $m_{H^+}$. For the limits shown in Fig.~\ref{fig:ATLASbounds} and discussed above, we have taken $m_{H_2}~=~m_{H^+}$= 130 GeV. In this case, no tree-level decay to mediators with masses satisfying present collider bounds is possible. The other case is $m_{H^+} = 300~\mathrm{GeV}$ which was used in some of our BR plots in Figure~\ref{fig:Xdecay}. But we did not use it in Fig.~\ref{fig:ATLASbounds} in deriving our limits on $\Lambda/\beta$. Therefore, our assumption of 100\% branching of $H^+\to W^+X$ is consistent.

Finally, it is worth noting that  for internal particle masses near $100$ GeV, one may be near the border of the region of validity of a pure effective theory treatment of the collider phenomenology. In principle, invoking an explicit model for generation of the operator coefficient and/or inclusion of a form factor would likely provide a more quantitatively realistic assessment. Similar considerations apply to the application of the Higgs effective theory in studies of Higgs boson observables (see, {\em e.g.}, Ref.~\cite{Dolan:2012rv} for a discussion in the context of SM di-Higgs production in association with an additional, high-$p_T$ jet). In the present instance, the ATLAS lepton jet reconstruction efficiency peaks in the vicinity of $p_T^X\sim 40$ GeV, while the masses of the intermediate $H^\pm/H_2$ and final state $W$-boson are not so large. Thus, we would expect at most a modest degradation of the signal strength in a more realistic, model-dependent analysis. Nonetheless, we consider our statements about the present LHC reach as indicative of the 8 TeV sensitivity rather than as quantitatively definitive.

\section{Outlook}\label{sec:conclusions}
Mixing between the dark U(1$)^\prime$ and SU(2$)_L$ gauge groups, mediated by the operator $\mathcal{O}_{WX}^{(5)}$, leads to a small mixing parameter $\epsilon$, whose magnitude is set by the scale ratio $v_\Sigma/\Lambda$ with an $\mathcal{O}(1)$ Wilson coefficient, $\beta$. The resulting collider phenomenology is quite distinctive, as $\mathcal{O}_{WX}^{(5)}$ may dominate  the production of final states containing $X$ bosons when $\Lambda/\beta\lsim 1$ TeV at both $\sqrt{s} = 8$ TeV and  $\sqrt{s} = 14$ TeV . Current ATLAS bounds, based on an inclusive search for pairs of lepton jets associated with displaced vertices, exclude $\Lambda/\beta $ up to about 600 GeV, depending on the value of $m_X$ and the triplet vev $v_\Sigma$. Looking to the future, the collection of additional data during Run II will extend the reach of the inclusive search. In the advent of a discovery, inclusion of additional search criteria associated with the final state vector boson(s) would allow one to distinguish this scenario from those associated with abelian kinetic mixing. An analysis of this possibility, along with the LHC sensitivity to other regions of the ($m_X$, $\epsilon$) plane, will appear in future work.

\acknowledgments

We thank Patrick Draper, Jesse Thaler, and Jiang-Hao Yu  for useful discussions. This work was supported in part by
U.S. Department of Energy contract DE-SC0011095 (G.O., T.P., and M.J.R.-M.). XGH was supported in part by MOE Academic Excellent Program (Grant No.~105R891505) and MOST of ROC (Grant No.~MOST~104-2112-M-002-015-MY3), and in part by NSFC of PRC (Grant No.~11575111). This work was also supported by Key Laboratory for Particle Physics, Astrophysics and Cosmology, Ministry of Education, and Shanghai Key Laboratory for Particle Physics and Cosmology (SKLPPC) (Grant No.~11DZ2260700). He also thanks Korea Institute for Advanced Study (KIAS) for their hospitality during the completion of this work. 
CA was supported in part by the National Science 
Foundation (ANT- 0937462, PHY-1306958, PHY-1505855, and PHY-1505858) and 
by the University of Wisconsin Research Committee with funds granted 
by the Wisconsin Alumni Research Foundation.

\appendix

%\section{Muon $g-2$ mass functions}\label{sec:AppendixB}
%The two mass functions that enter the $g-2$ formula in \eq{eq:gminus2formula} are
%\begin{eqnarray}
%G_1(x)=\frac{f_1(x)-f_2(x)}{2},\qquad G_2(x)=f_1(x)+f_2(x),
%\end{eqnarray}
%where
%\begin{eqnarray}
%f_1(x)=\int_0^1\,\d z\,\frac{z(1-z)^2}{z+(1-z)^2\, x},\qquad f_2(x)=\int_0^1\,\d z\,\frac{z(1-z)(3+z)+2(1-z)^3\,x}{z+(1-z)^2\, x}\,.
%\end{eqnarray}
%Performing these integrals we for convenience provide analytical formulas for these functions
%\begin{eqnarray}
%&&G_1(x)=\frac{-x(1-x)(2-x)-[1-x(3-x)]\ln x+\frac{1-5x(1-x)}{\sqrt{1-4x}}\ln\frac{4x}{(1+\sqrt{1-4x})^2}}{2x^3},\\
%&&G_2(x)=\frac{-2x+x^2-(1-x)\ln x+\frac{1-3x}{\sqrt{1-4x}}\ln\left[\frac{4x}{(1+\sqrt{1-4x})^2}\right]}{x^2}\,.
%\end{eqnarray}

\section{Feynman rules relevant for collider signatures}\label{sec:AppendixC}
Feynman rules of interactions between the dark bosons, leptons, gauge bosons, charged and neutral Higgs bosons are listed in the table below:
\begin{center}
  \begin{tabular}{ |l | c | r }
    \hline
    \text{Interaction} &  \text{Feynman rule}  \\ \hline
     $ X l^+ l^-$ & $i e\left(\epsilon_0-\frac{\beta v_{\Delta}s_W}{\Lambda}\right)$ \\ \hline
   $ W^{\pm}H^{\mp}X$ & $\frac{i\beta}{\Lambda}\left(g^{\mu\nu}pp'-p^{\nu}p '^{\mu}\right)c_{\mp}$ \\ \hline
   $ Z H_1 X$ & $\frac{i\beta}{\Lambda}\left(g^{\mu\nu}pp'-p^{\nu}p '^{\mu}\right)c_{W} s_0$ \\ \hline
   $ Z H_2 X$ & $\frac{i\beta}{\Lambda}\left(g^{\mu\nu}pp'-p^{\nu}p '^{\mu}\right)c_W c_0$ \\ \hline
   $ A H_1 X$ & $\frac{i\beta}{\Lambda}\left(g^{\mu\nu}pp'-p^{\nu}p '^{\mu}\right)s_W s_0$ \\ \hline
   $ A H_2 X$ & $\frac{i\beta}{\Lambda}\left(g^{\mu\nu}pp'-p^{\nu}p '^{\mu}\right)s_W c_0$ \\ \hline
   \hline
     $ W^+_{\mu}(p_1)W^-_{\nu}(p_2)H_1 X_{\alpha}(p_3)$ & $\frac{i\beta\,g}{\Lambda}\left(p_3^{\mu}g^{\nu\alpha}-p_3^{\nu}g^{\mu\alpha}\right) s_0$ \\ \hline
     $ W^+_{\mu}(p_1)W^-_{\nu}(p_2)H_2 X_{\alpha}(p_3)$ & $\frac{i\beta\,g}{\Lambda}\left(p_3^{\mu}g^{\nu\alpha}-p_3^{\nu}g^{\mu\alpha}\right) c_0$ \\ \hline
     $ W^{\pm}_{\mu}(p_1) Z_{\nu}(p_2)\,H^{\mp} X_{\alpha}(p_3)$ & $\mp\frac{i\beta\,g}{\Lambda}\left(p_3^{\mu}g^{\nu\alpha}-p_3^{\nu}g^{\mu\alpha}\right)c_W\,c_{\mp}$ \\ \hline
     $ W^{\pm}_{\mu}(p_1) A_{\nu}(p_2)\,H^{\mp} X_{\alpha}(p_3)$ &$\mp\frac{i\beta\,g}{\Lambda}\left(p_3^{\mu}g^{\nu\alpha}-p_3^{\nu}g^{\mu\alpha}\right)s_W\,c_{\mp}$ \\ \hline      
  \end{tabular}
\end{center}
 Feynman rules entering in the vertices of the graphs in Figure~\ref{fig:LHCsignatures}. Where $c_{\mp}\equiv\cos\theta_{\mp}$ and $c_0\equiv\cos\theta_0$ are as defined in Ref.~\cite{FileviezPerez:2008bj}.

%\paragraph{Note added.} This is also a good position for notes added
%after the paper has been written.

% The bibliography will probably be heavily edited during typesetting.
% We'll parse it and, using the arxiv number or the journal data, will
% query inspire, trying to verify the data (this will probalby spot
% eventual typos) and retrive the document DOI and eventual errata.
% We however suggest to always provide author, title and journal data:
% in short all the informations that clearly identify a document.

\bibliographystyle{h-physrev}
   \let\oldnewblock=\newblock
    \newcommand\dispatcholdnewblock[1]{\oldnewblock{#1}}
    \renewcommand\newblock{\spaceskip=0.3emplus0.3emminus0.2em\relax
                           \xspaceskip=0.3emplus0.6emminus0.1em\relax
                           \hskip0ptplus0.5emminus0.2em\relax
                           {\catcode`\.=\active
                           \expandafter}\dispatcholdnewblock}
\bibliography{bibliography}

% Please avoid comments such as "For a review'', "For some examples",
% "and references therein" or move them in the text. In general,
% please leave only references in the bibliography and move all
% accessory text in footnotes.

% Also, please have only one work for each \bibitem.

\end{document}